\documentclass[12pt,preprint]{aastex}

\usepackage{natbib}
\usepackage{aas_macros}
\newcommand{\dfrac}[2]{{\displaystyle \frac{#1}{#2}}  }

\newcommand{\eqref}[1]{(\ref{#1})}

\def\lesssim{\mathrel{\hbox{\rlap{\hbox{\lower4pt\hbox{$\sim$}}}\hbox{$<$}}}}
\def\gtrsim{\mathrel{\hbox{\rlap{\hbox{\lower4pt\hbox{$\sim$}}}\hbox{$>$}}}}

\slugcomment{\today}
\shorttitle{}
\shortauthors{K\.D\. Kanagawa et al\.}

\begin{document}
\title{Mass Estimates of a Giant Planet in a Protoplanetary Disk from
the Gap Structures}

\author{Kazuhiro D. KANAGAWA\altaffilmark{1}, 
Takayuki MUTO\altaffilmark{2}, 
Hidekazu TANAKA\altaffilmark{1}, 
Takayuki TANIGAWA\altaffilmark{3}, 
Taku TAKEUCHI\altaffilmark{4},
Takashi TSUKAGOSHI\altaffilmark{5},
Munetake MOMOSE\altaffilmark{5},
}
\email{kanagawa@lowtem.hokudai.ac.jp}
\altaffiltext{1}{Institute of Low Temperature Science, Hokkaido
University, Sapporo 060-0819, Japan} 
\altaffiltext{2}{Division of Liberal Arts, Kogakuin University, 1-24-2,
Nishi-Shinjuku, Shinjuku-ku, Tokyo, 163-8677, Japan} 
\altaffiltext{3}{School of Medicine, University of Occupational and
Environmental Health, Yahatanishi-ku, Kitakyushu, Fukuoka 807-8555,
Japan} 
\altaffiltext{4}{Department of Earth and Planetary Sciences, Tokyo
Institute of Technology, Meguro-ku, Tokyo 152-8551, Japan} 
\altaffiltext{5}{College of Science, Ibaraki University,  
2-1-1, Bunkyo, Mito, Ibaraki 310-851, Japan} 

\begin{abstract}
A giant planet embedded in a protoplanetary disk forms a gap.
An analytic relationship among the gap depth, planet mass $M_{p}$, disk aspect ratio $h_p$, and viscosity $\alpha$ has been found recently, and the gap depth can be written in terms of a single parameter $K= (M_{p}/M_{\ast})^2 h_p^{-5} \alpha^{-1}$.
We discuss how observed gap features can be used 
to constrain the disk and/or planet parameters based on the analytic formula for the gap depth.
The constraint on the disk aspect ratio is critical in determining the
planet mass so the combination of the observations of the temperature
and the image can provide a constraint on the planet mass.
We apply the formula for the gap depth to observations of HL~Tau and HD~169142.
In the case of HL~Tau, we propose that a planet with $\gtrsim 0.3$ is responsible for the observed gap at $30$~AU from the central star based on the estimate that the gap depth is $\lesssim 1/3$.
In the case of HD~169142, the planet mass that causes the gap structure recently found by VLA is $\gtrsim 0.4 M_J$.
We also argue that the spiral structure, if observed, can be used to estimate the lower limit of the disk aspect ratio and the planet mass.
\end{abstract}

\keywords{protoplanetary disks --- planet-disk interactions ---
stars:individual (HD 169142, HL Tau)}

\section{Introduction}
\label{sec:intro}

High resolution imaging observations of protoplanetary disks 
have revealed the presence of complex morphological
structures in disks such as
dips~\citep[e.g.,][]{Hashimoto2012}, 
gap~\citep[e.g.,][]{Debes2013}, or
spirals~\citep[e.g.,][]{Muto2012,Grady2013} in near infrared (NIR)
scattered light.
At submillimeter wavelengths, ALMA has revealed the existence of
non-axisymmetric structures in its early science operations 
\citep[e.g.,][]{Casassus2013,Fukagawa2013,vanderMarel2013,Perez2014} 
and now multiple ring-like structures are discovered in 2014
Long Baseline Campaign~\citep{ALMA_HLTau}. 
It is now possible to obtain high spatial resolution
data with the beam size less than $0\farcs04$ of multiple wavelengths.

One possible origin of the detail structures in protoplanetary
disks is the disk--planet interaction.
A giant planet creates a gap and spirals in the disk along its orbit, 
which are observable with high angular resolution observations.
It is therefore important to understand what we can learn about the
physical properties of the disk and planet once we can observe such
structures.  

In this letter, we present an estimation method of the planet mass from the gap depth which is potentially an observable quantity. 
The gap depth is related to the planet mass, the viscosity and the disk scale height (or disk temperature) \citep[hereafter, K15]{Duffell_MacFadyen2013,Fung_Shi_Chiang2014,Kanagawa2015}.
Using this relation, we can constrain the mass of a planet, if exists, by the combination of the disk image and observations of disk temperature and viscosity.
The disk temperature can be obtained from the observation of optically thick gas/dust emissions.
Alternatively, the spiral structure associated with the gap can be used to constrain the lower limit of the disk scale height. If the gap depth and the spiral structure are both observed, it enables us to derive the lower limit of the planet mass.

In Section \ref{sec:gapmodel}, we present an analytic formula that
describes the gap depth in terms of the physical parameters of the disk and the
planet.   
In Section \ref{sec:observations}, we apply
the gap depth formula to the observations of  HL~Tau by ALMA \citep{ALMA_HLTau} and 
HD~169142 by VLA \citep{Osorio2014}. 
In Section \ref{sec:spiralconstraint}, we discuss the constraints on the
disk aspect ratio derived from spiral structure.
Section \ref{sec:summary} is for summary and discussion.

\section{Gap model}
\label{sec:gapmodel}

In this section, we describe an analytic formula that relates the
gap depth, planet mass and disk properties.
A giant planet embedded in a protoplanetary disk creates a gap \citep[e.g.,][]{Lin_Papaloizou1979,Goldreich_Tremaine1980}, 
which has been investigated by many authors \citep[e.g.,][]{Artymowicz_Lubow1994,Kley1999,Varnire_Quillen_Frank2004,Crida_Morbidelli_Masset2006}. 
Recent hydrodynamic simulations by \cite{Duffell_MacFadyen2013} and \cite{Fung_Shi_Chiang2014} 
give an empirical formula for the relationship between planet mass and
 the gap depth,  
which is derived analytically by considering the angular momentum
exchange processes between the disk and the planet 
\citep[][and K15]{Fung_Shi_Chiang2014}.
We briefly explain a more accurate derivation given by K15, below.

We assume a steady, axisymmetric, non--self--gravitating, geometrically thin disk.
In a viscous accretion disk, the radial angular momentum flux $F_{J}$ is given by the combination of
the fluxes carried by advection and viscosity \citep{Lynden-Bell_Pringle1974},
\begin{equation}
	F_{J}(R) = R^2\Omega F_{M} 
	 - 2\pi R^3 \nu \Sigma \frac{\partial \Omega}{\partial R},
	\label{eq:amf}
\end{equation}
where $R$, $\Omega$, $F_{M}$, $\nu$, and $\Sigma$ denote the radial distance from the central star, the angular velocity of the disk, the mass flux, the kinetic viscosity, and the surface density, respectively. 
Adopting the $\alpha$--prescription of kinematic viscosity \citep{Shakura_Sunyaev1973}, we write $\nu=\alpha h^2 R^2 \Omega$, 
where $h=c/(R \Omega)$ denotes the disk aspect ratio and $c$ is the sound speed, which are dependent on the disk temperature.

Because of a strong gravitational torque exerted by a planet on the disk, a gap opens along the planet orbit.
We assume a sufficiently wide gap so that the torque is exerted mainly within a flat gap bottom with a constant surface density  $\Sigma_p$.
Then, using the WKB torque formula with a cutoff, the one--sided Lindblad torque, $T_{\rm LB}$, which is exerted by the planet on the outer disk, is approximately given by \citep[see][and K15]{Lin_Papaloizou1979}
\begin{eqnarray}
 T_{\rm LB} &=& 
 \displaystyle{\int}^{\infty}_{R_p+\Delta} 0.80 \pi
  \left( \frac{M_{p}}{M_{\ast}} \right)^{2} 
  R_p^3 \Omega_p^2 \Sigma_p \left( \frac{R_p}{R-R_p} \right)^{4} dR \nonumber \\
 &=& 0.12 \pi \left( \frac{M_{p}}{M_{\ast}} \right)^2 h_p^{-3} R_p^4 \Omega_{\rm p}^2 \Sigma_p,
     	\label{eq:outerLB}
\end{eqnarray}
where $M_{p}$ and $M_{\ast}$ denotes the masses of the planet and the central star, respectively, and the subscript $p$ denotes the values at the planet orbital radius $R_p$. 
In the above, we set the torque cutoff length $\Delta=1.3h_p R_p$ to match the one--sided torque with the results by detailed linear analyses of disk--planet interaction
\citep{Takeuchi_Miyama1998,Tanaka_Takeuchi_Ward2002,Muto_Inutsuka2009}.
The above assumption that the planetary torque is exerted mainly within the gap can be valid when the gap width is much larger than the cutoff length.

\begin{figure}
  \centering
  \resizebox{0.47\textwidth}{!}{\includegraphics{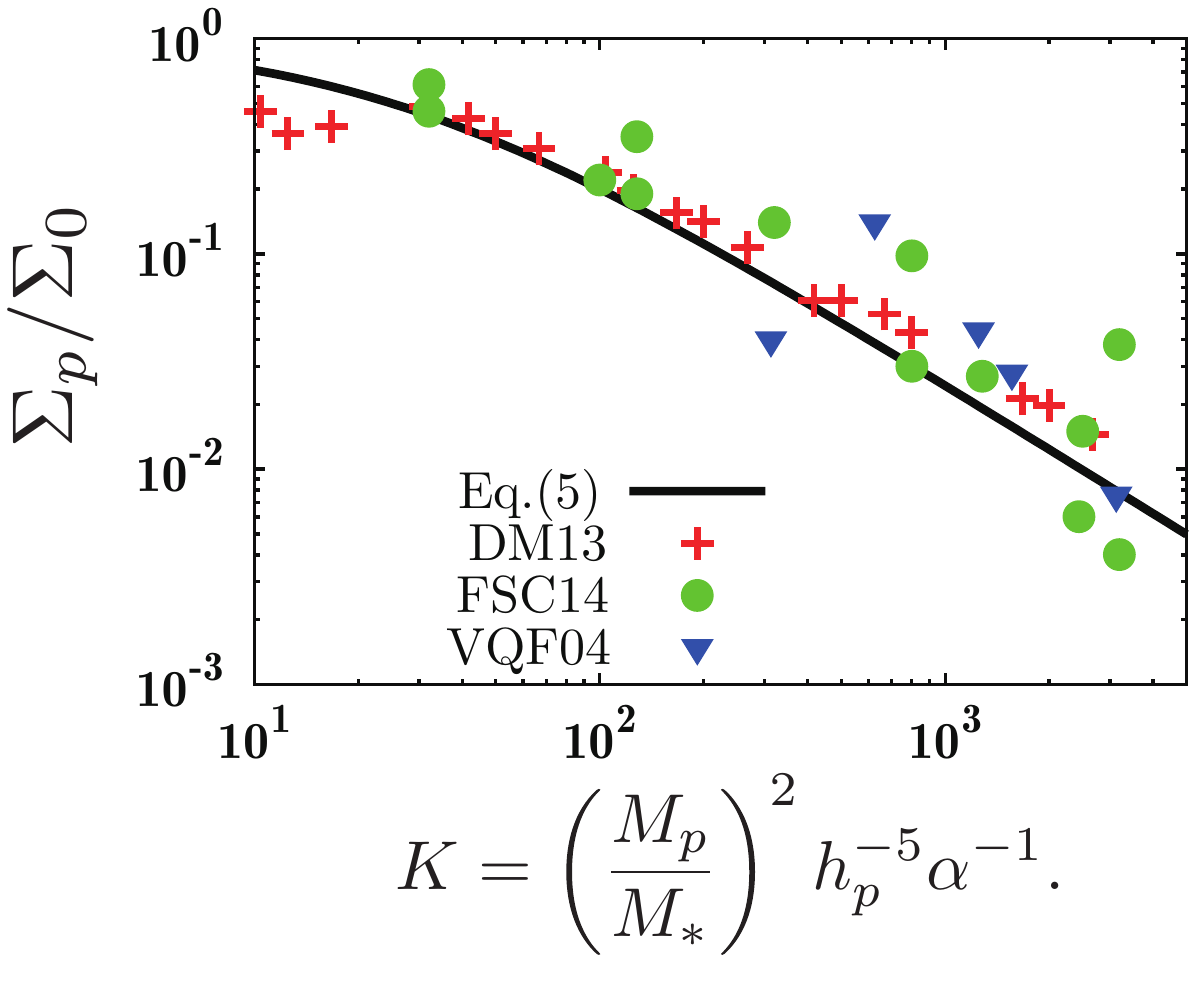}}
  \caption{
  The gap depth as a function of the parameter $K$.
  The solid line denotes Equation~(\ref{eq:gap_depth}).
  The crosses, circles and triangles represent the gep depth obtained by
  hydrodynamic simulations done by 
  \citet{Duffell_MacFadyen2013}, 
  \cite{Fung_Shi_Chiang2014} and 
  \cite{Varnire_Quillen_Frank2004}, respectively.
  (entire runs of $K<8000$ in these papers are shown)
  }
	\label{fig:gapdepth}
 \end{figure}

In steady state, the mass flux is constant and the angular momentum flux radially changes due to the planetary torque.
Thus, the angular momentum flux at the outside of the gap, $R_e (>R_p)$ is equal to the sum of the angular momentum flux at $R_p$ and the one--sided torque $T_{\rm LB}$,
\begin{equation}
	F_J(R_{e}) = F_{J}(R_{p})+T_{{\rm LB}}.
	\label{eq:amf_planet}
\end{equation}
At the outside of the gap, the surface density is the unperturbed value $\Sigma_0$.
The gap width is much smaller than $R_p$ though the width is longer than the cutoff length.
Then we can set $R_e\simeq R_p$, $h_e\simeq h_p$, and $\Omega_{e} \simeq \Omega_p$, and therefor $R_e^2\Omega_e F_{\rm M} \simeq R_p^2 \Omega_p F_{\rm M}$.
Under these approximations, substitution of Equations~\eqref{eq:amf} and \eqref{eq:outerLB} into \eqref{eq:amf_planet} yields
\begin{equation}
  3\pi \alpha h_{p}^{2} R_{p}^{4} \Omega_{p}^{2} \Sigma_{0}
  =  
 3\pi \alpha h_{p}^{2} R_{p}^{4} \Omega_{p}^{2} \Sigma_{p}
  + 0.12 \pi \left( \dfrac{M_{p}}{M_{\ast}} \right)^2 h_{p}^{-3} R_{p}^{4} \Omega_{p}^2 \Sigma_p.
  \label{eq:amf_balance}
\end{equation}
From Equation~\eqref{eq:amf_balance}, we obtain the gap depth $\Sigma_{p}/\Sigma_0$, as
\begin{equation}
	\frac{\Sigma_p}{\Sigma_0} = \frac{1}{1+0.04 K},
	\label{eq:gap_depth}
\end{equation}
where
\begin{equation}
	K = \left( \frac{M_{p}}{M_{\ast}} \right)^2 
	 h_{p}^{-5} \alpha^{-1},
	\label{eq:K}
\end{equation}
(see K15 for detail derivation).
Although we adopt a simple expression for the one--sided torque, Equation~\eqref{eq:gap_depth} gives a good agreement with the results of hydrodynamic simulations as shown in Figure~\ref{fig:gapdepth}.

Equation \eqref{eq:gap_depth} would be valid for $K\lesssim 10^{4}$.
If $K$ is appreciably large, 
the gap even becomes eccentric and non--steady, which breaks down our simple estimate \citep{Kley_Dirksen2006}.
In the eccentric state, the gap is
shallower than the prediction by Equation \eqref{eq:gap_depth}
\citep{Fung_Shi_Chiang2014}.  

\begin{figure}
 \centering
 \resizebox{0.94\textwidth}{!}{\includegraphics{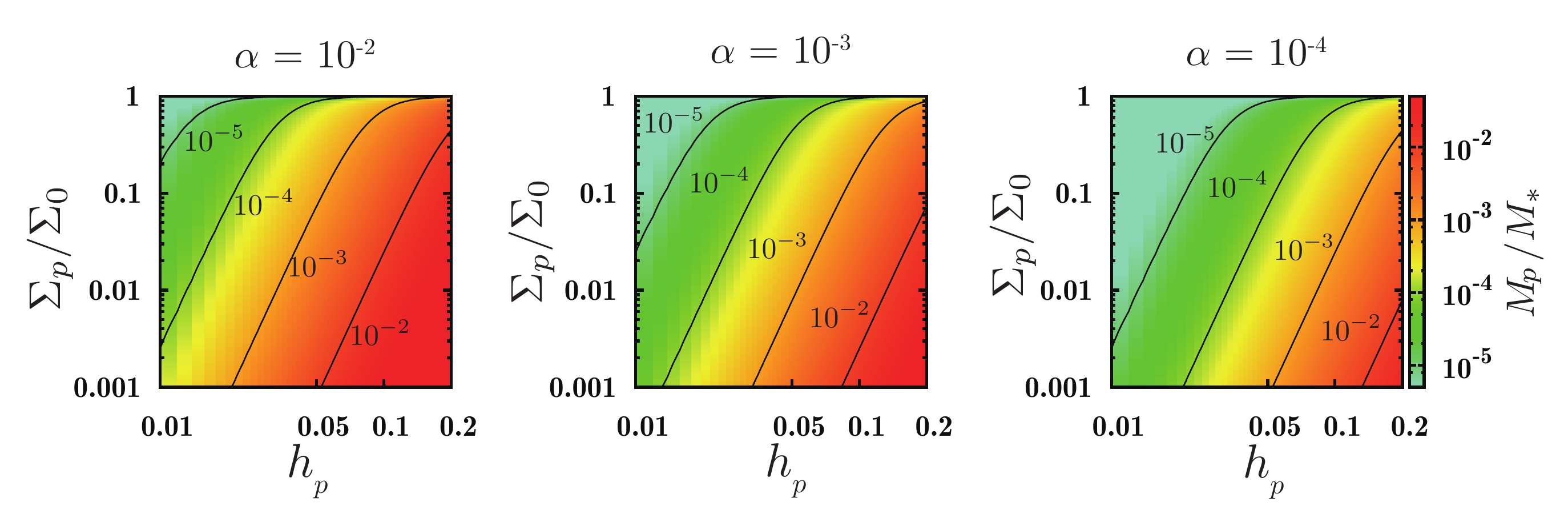}}
 \caption{Planet mass as a function of the gap depth and the disk aspect
 ratio.  The relationship between these values are given in Equation
 \eqref{eq:gap_depth_Mpl} and three panels correspond to the cases with 
 $\alpha=10^{-2}$ (left), $10^{-3}$ (center) and $10^{-4}$ (right).
 }
 \label{fig:param_contour}
\end{figure}

As seen in Equation \eqref{eq:gap_depth}, 
it is related to the planet mass, 
the viscosity and the aspect ratio.
Solving Equation \eqref{eq:gap_depth} for $M_{p}/M_{\ast}$, 
we obtain
\begin{equation}
 \frac{M_{p}}{M_{\ast}} = 5 \times 10^{-4} \left( 
                  \frac{1}{\Sigma_{p}/\Sigma_{0}} -1
			      \right)^{1/2} 
 \left( \dfrac{h_{p}}{0.1} \right)^{5/2} 
 \left( \dfrac{\alpha}{10^{-3}} \right)^{1/2}.
  \label{eq:gap_depth_Mpl}
\end{equation}
Figure~\ref{fig:param_contour} displays the planet mass given by Equation~\eqref{eq:gap_depth_Mpl} as a function of
the gap depth $\Sigma_p/\Sigma_0$ and the disk aspect ratio for several cases of $\alpha$.  
The dependence of the planet mass on $h_p$ is strong, though $\alpha$--dependence is relatively weak.
The planet mass varies from $\sim 10^{-5}$ to $\sim 10^{-2}$ if $h_p$ changes from $0.01$ to $0.2$ for $\Sigma_p/\Sigma_0 =1/10$ and $\alpha=10^{-3}$.
Therefore, the constraint on $h_p$, or equivalently disk temperature, is important in estimating the planet mass.
In current observations, the constraints of the disk temperature can be given, for example, by emissions from optically thick gas (or sometimes dust) as discussed in Section~\ref{sec:observations}.
As discussed in Section~\ref{sec:spiralconstraint}, we also show that the spiral structure can constrain the lower limit of the disk aspect ratio.

In application of our model to an observed gap, we assume that the distribution of dust is similar to the gas.
Because of the dust filtration accompanied with the gap, however, the dust distribution can be significantly deviated from the gas , as reported by many theoretical studies \citep[e.g.,][]{Paardekooper2004,Rice2006,Ward2009,Zhu2012}.
The millimeter--sized or larger dust particles strongly suffer the dust filtration.
They are piled up at the outer edge of the gap and have a much deeper gap than the gas.
Smaller particles with radii of $\lesssim 0.1$mm, on the other hand, are not much affected by the dust filtration and their distributions are similar to the gas for a gap created by a Jupiter mass planet \citep[see Figure~{3} of][]{Zhu2012}.
Hence the above assumption in our model is valid for the case of such small dust particles.
Also note that the critical dust size of filtration depends on the gas density and the turbulent viscosity.
\cite{Zhu2012} did not take into account an additional diffusion of dust particles by the turbulence due to the Rossby wave instability and the Rayleigh unstable region \citep[e.g.,][]{Zhu2014,Kanagawa2015}, which occur at a deep gap.
This additional diffusion can further increase the critical dust size for filtration.
In application to an observed gap, it is also important to judge whether the filtration effect strongly alters the dust distribution at the gap or not.
Such a judgement can be possible, by checking the dust pilling--up at the outer edge of the gap (see Section~\ref{sec:observations}).

\section{Application to Observations}
\label{sec:observations}
In the actual observations of protoplanetary disks to date, 
there are several cases where the gap-like structures are
observed.   
In this section, we consider two examples: HL~Tau and HD~169142.

HL~Tau was observed during the 2014 Long
Baseline Campaign of ALMA \citep{ALMA_HLTau}.  The observations show
multiple rings but we focus on the ring residing at $\sim 30$~AU, 
which is clearly seen and reasonably resolved.

\begin{figure}
 \centering
 \resizebox{0.94\textwidth}{!}{\includegraphics{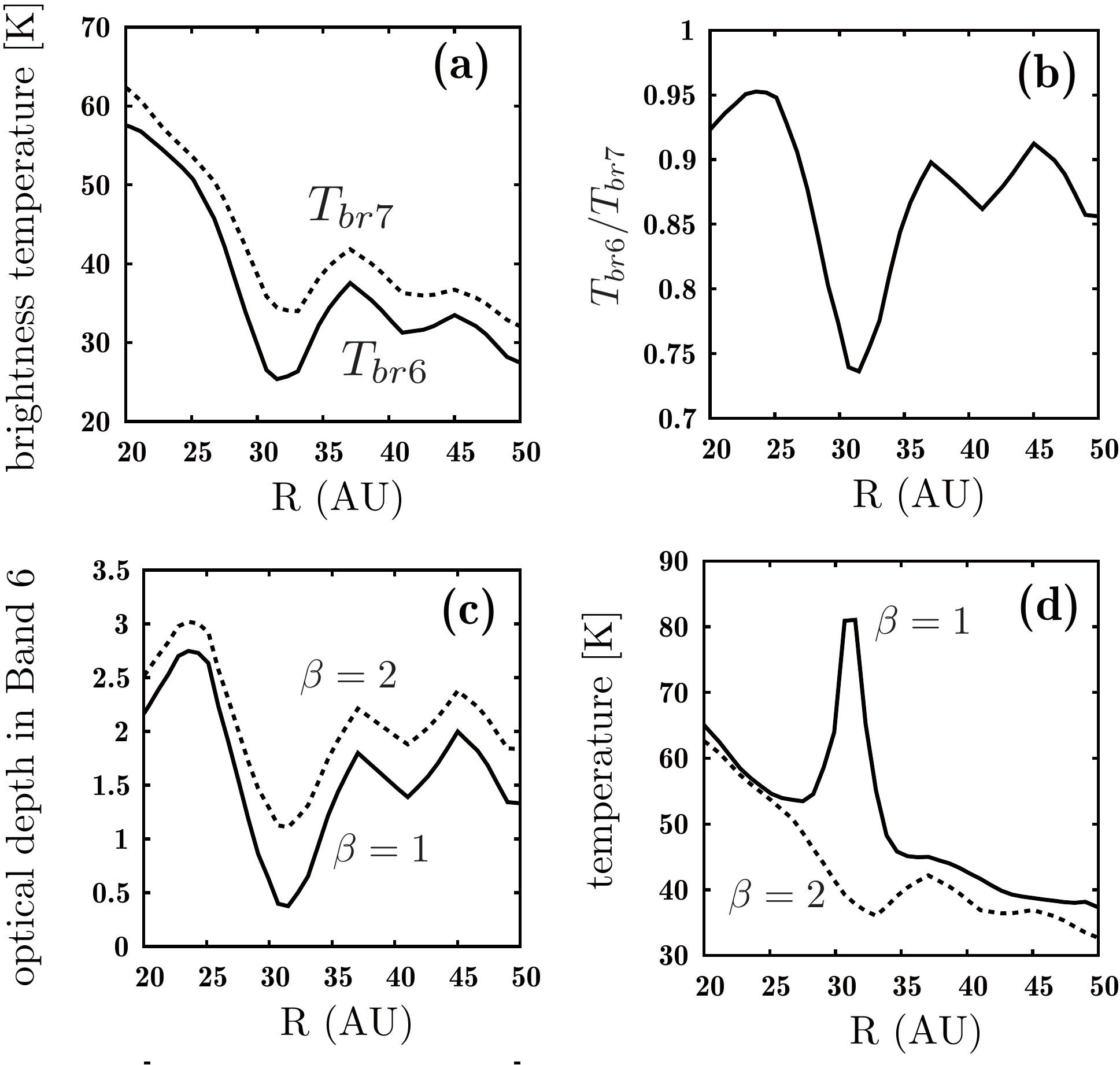}}
 \caption{
 {(a)}: Observed radial profile of the brightness
 temperature of dust continuum emission in HL~Tau disk along
 the major axis.  The data is averaged 
 over PA=$135^{\circ}\pm 5^{\circ}$.  Solid
 line shows the Band 6 data and the dashed line shows the Band 7 data.
 {(b)}: The ratio of brightness temperature in Band 6 and 7.
 {(c)}: The radial profiles of the optical depth.
 {(d)}: The temperature.
 The derivations of (c) and (d) are described in Section~\ref{sec:observations}.  
 }
 \label{fig:prof_results}
\end{figure}

We use the FITS image of the HL~Tau that is publicly available from the ALMA archive site. 
We use the data of dust continuum emission in Band 6 and {7} in our analyses, and assume the distance of $140$~pc.
Figure \ref{fig:prof_results}--(a) and (b) show the radial profiles of the brightness temperature in Band 6 and 7 between
$20~\mathrm{AU}$ -- $50~\mathrm{AU}$ averaged over the position angles of 
$130^{\circ}<\mathrm{PA}<140^{\circ}$, which is close to the major axis
of the disk.  
We use the position of the central brightest pixel as the location of
the disk center.  
Since the brightness temperatures in Band 6 and 7 are similar at 
$R \lesssim 25~\mathrm{AU}$, the dust emission is likely to be optically thick.
Therefore, we use the ratio of the brightness temperature in the
two bands to estimate the actual gap depth.  
The brightness temperatures in Band 6 and 7, $T_{br6}$ and $T_{br7}$, are
given by
\begin{equation}
 T_{br6} = T \left( 1-e^{-\tau_{6}} \right)
  \label{eq:tau6}
\end{equation}
and
\begin{equation}
 T_{br7} = T \left( 1-e^{-\tau_{7}} \right)
  \label{eq:tau7}
\end{equation}
where $T$ is the disk temperature, $\tau_{6}$ is the optical depth in
Band 6 and $\tau_{7}$ is that in Band 7.  The optical depths of the Band
6 and 7 are related by $\tau_7 = \tau_6 (\nu_7/\nu_6)^{\beta}$, where
$\nu_6=233~\mathrm{GHz}$ is the central frequency at Band 6,  
$\nu_7=344~\mathrm{GHz}$ is that at Band 7, and $\beta$ is the opacity
index.  We can derive $\tau_{6}$ and $T$ by using Equations
\eqref{eq:tau6} and \eqref{eq:tau7} if we assume $\beta$.  
Figure \ref{fig:prof_results}--(c) shows the radial profiles of $\tau_6$ in the
cases of $\beta=1$ and $\beta=2$.  By measuring the contrast of the
maximum of $\tau$ at $R\sim25~\mathrm{AU}$ and the minimum at
$R\sim30~\mathrm{AU}$, we estimate the depth of the gap to be 
$\Sigma_p/\Sigma_0 \sim \tau_{\max}/\tau_{\min} \sim 1/3-1/7$.

Figure \ref{fig:prof_results}--(d) also shows the profile of $T$ in the cases of
$\beta=1$ and $2$.  The temperature in the gap region strongly depends
on the assumption of $\beta$, but that outside the gap is relatively
insensitive to it.  If we use $T=55~K$ at
$R=25~\mathrm{AU}$, and assume that 
the sound speed is $c=1~\mathrm{km/s} (T/300~\mathrm{K})^{1/2}$ and 
that the central star mass is $1~M_{\odot}$, we derive 
$h \sim 0.07$ as an estimate
of the disk aspect ratio.
Adopting Equation \eqref{eq:gap_depth_Mpl}, we derive
\begin{equation}
 \dfrac{M_{p}}{M_{\ast}} = 3 \ \mbox{--} \ 5 \times 10^{-4} 
  \left( \dfrac{h_p}{0.07} \right)^{5/2} 
  \left( \dfrac{\alpha}{10^{-3}} \right)^{1/2},
\end{equation}
if this gap is created by a planet.  
Note that the error in the estimate of the stellar mass also increases the error in the planet-star mass-ratio through $h_p \propto c/(R_p\Omega_p) \propto M_{\ast}^{-1/2}$.
In the above estimate, it is assumed that dust particles are gas have simular distributions.
This assumption is vaild when dust particles are too samll to be much affected by the dust filtration, as mentioned in the end of Section~\ref{sec:gapmodel}.
We can check the validity of the assumption, using the observed gap structure.
According to the result by \cite{Zhu2012}, the dust surface density at the outer edge of the gap is much larger than that at the inner edge when the dust filtration is effective.
For the gap at $30$ AU of HL~Tau, on the other hand, the distribution of the dust optical depth (see Figure~\ref{fig:prof_results}--[c]) does not show any enhance at the outer edge.
Hence, we can judge that dust filtration is not so effective in this case.

The second example is the disk around HD~169142.
Recent observations by \cite{Osorio2014} of 7~mm emission shows that
there may exist a gap at $\sim 50$~AU from the central star.
The observations also found a knot at $50$~AU, which could be regarded as compact source with the mass of $\sim 0.6 M_{J}$.
It is difficult to estimate the gap depth from their observation since the beam size is
relatively large, but the radial profile given in the Figure~2 of
\citet{Osorio2014} suggests that the gap may be as shallow as 
$\Sigma_p/\Sigma_0 \simeq 0.5$.  
We consider this as the upper limit of the gap depth 
since the model with complete gap
(i.e., $\Sigma_p/\Sigma_0 = 0$) seems to be consistent with
observations.  
If we adopt the disk aspect ratio of 
$\sim 0.1$ and $\alpha=5\times10^{-4}$, as indicated by
\citet{Osorio2014}, we obtain the lower limit of the planet mass as 
 $\sim 0.4~M_J$ from Equation \eqref{eq:gap_depth_Mpl}.

\section{The Constraint on Disk Temperature from Spiral Structures}
\label{sec:spiralconstraint}

We have seen that the constraint on disk temperature, or disk aspect ratio,
is important in inferring the planet mass from the gap structure.  
As discussed above, the disk temperature is obtained from the disk emissions.
Here, we discuss an alternative constraint on disk aspect ratio coming from a spiral density wave, which is a natural consequence of disk--planet interaction.
The spiral structures have been observed in infrared observations so far \citep[e.g.,][]{Muto2012,Grady2013,Currie2014}.
We expect such features may be observed at sub--millimeter wavelengths in near future when better sensitivity at higher spatial resolution is available.

This shape of the spiral density wave induced by disk--planet interaction is related to the disk aspect ratio (or temperature) and is given in an analytic form~\citep{Rafikov2002,Muto2012}.
The shape of spiral waves excited by a planet is determined by four parameters, $R_c, \theta_0, \delta$ and $h_p$ \citep[see eq.~{[1]}][]{Muto2012}.
The launching location of spiral waves (very close to the planet location) is denoted as ($R_c,\theta_0$).
The parameter $\delta$ denotes the profile of sound speed $c \propto R^{-\delta}$.
The disk aspect ratio $h$ changes as $h = h_p \left( R/R_p \right)^{-\delta+1/2}$.

\begin{figure}
 \centering
 \resizebox{0.94\textwidth}{!}{\includegraphics{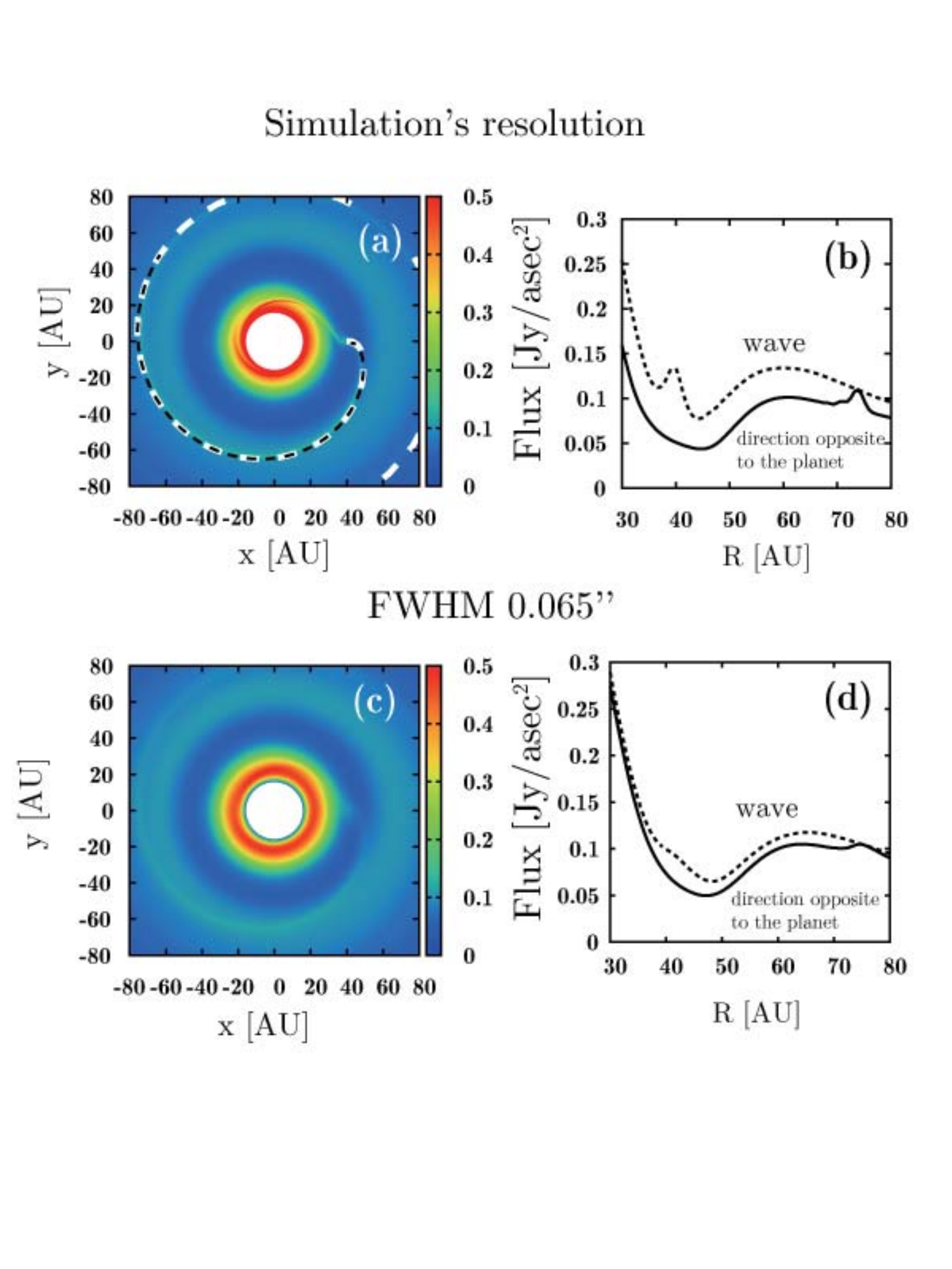}}
 \caption{
 {(a)}: The gap and spiral structure caused by a Jupiter mass planet in
 the disk with $\alpha=10^{-3}$, $h_p=0.1$ and $\delta=0.25$.  The black
 dashed line indicates the best-fit curve for the spiral shape
 and the white dashed line indicates the spiral used for fitting.
 {(b)}: The radial cuts of emission along the spiral (dashed) and at a direction opposite to the planet (solid).
 {(c)} and {(d)}: The same as (a) and (b), but the emission is convolved with a circular Gaussian beam with an FWHM of 0.065'' (the distance to the disk is set as $150$ pc).
 }
 \label{fig:wavefit}
\end{figure}
\begin{figure}
 \centering
 \resizebox{0.94\textwidth}{!}{\includegraphics{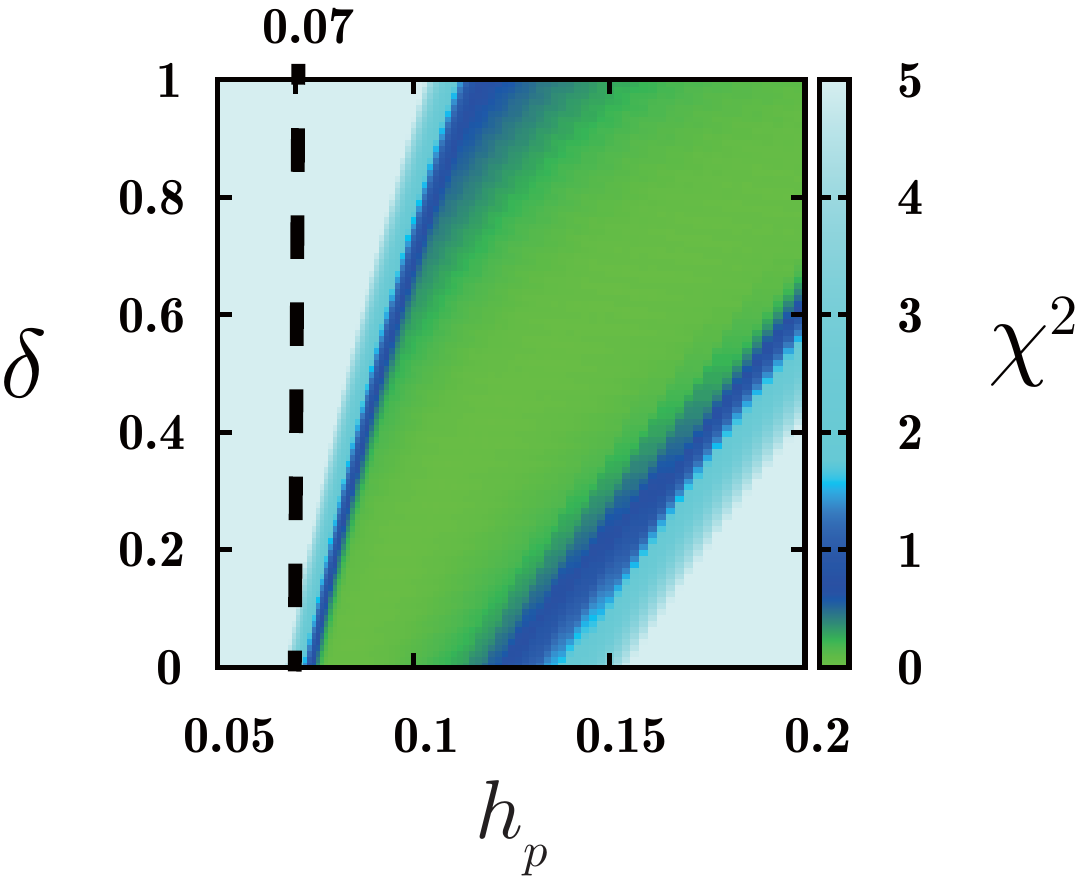}}
 \caption{The $\chi^2$-map of the fitting to the spiral structure.}
 \label{fig:paramfit}
\end{figure}

For demonstration, we have performed two--dimensional hydrodynamic
calculations using FARGO~\citep{Masset2000} 
to obtain the gap and spiral structures caused by
a Jupiter mass planet at $R_p=40~\mathrm{AU}$ in a disk with
$\alpha=10^{-3}$, $h_p=0.1$, and $\delta=0.25$.
Figure~\ref{fig:wavefit}-(a) and (b) are the optically thin dust thermal emission at 230GHz obtained by the simulation ($\kappa$ is assumed to be $2\times 10^{-2}$~cm$^2$/g per unit gas mass and the dust and gas are assumed to be well--mixed), and the radial cuts of the emission along the wave and in the direction opposite to the planet.
Figure~\ref{fig:wavefit}-(c) and (d) are the same as (a) and (b), but the emission is convolved with a circular Gaussian beam with an FWHM of 0.065'' (the distance to the disk is set as $150$ pc).
Even in the convoluted image, the emission of wave is about $10\%$ larger than the emission from non--wave region near the planet.
Such spiral feature would be detectable with future ALMA or TMT observations.
We fit the spiral structure, and the resulting $\chi^2$--map (Figure~\ref{fig:paramfit}) indicates that we can constrain the lower limit of $h_p$ to be $\sim 0.07$.
The gap depth can be estimated by $\Sigma_p/\Sigma_0 \sim 1/5$ from Figure~\ref{fig:wavefit}.
Therefore, from Equation \eqref{eq:gap_depth_Mpl}, 
we derive the lower limit of the planet mass as 
$M_p/M_{\ast} \geq 4\times 10^{-4} (\alpha/10^{-3})^{1/2}$.
This lower limit actually coincides with the actual planet mass within a
factor of $\sim 2.5$ in this particular case.

\section{Summary and Caveats}
\label{sec:summary}

In this letter, we have shown an analytic derivation of the depth of the gap created by a planet embedded in a protoplanetary disk.  
Equation \eqref{eq:gap_depth_Mpl} gives the relationship between the gap
depth, planet mass, disk temperature (aspect ratio) and viscosity.  
Since the gap depth strongly depends on the disk aspect ratio, the constraints on the disk temperature is critical
in estimating the planet mass from the gap structure.
The disk temperature is obtained by the emissions from optically thick gas/dust, as shown in Section~\ref{sec:observations}.
We have also argued that the spiral wave can constraint the disk temperature in the observation of optically thin emission.
Our model assumed that dust particles and gas have similar distributions.
This assumption is vaild when dust particles are too small to be much affected by the dust filtration.
The assumption would be verified if the dust piling--up is not observed at the outer edge of the gap.

We have applied Equation~\eqref{eq:gap_depth_Mpl} to the dust continuum
observations of  HL~Tau and HD~169142 and derived the estimates of the
mass of the planet, 
which could be responsible for the observed gaps (at 30~AU of HL~Tau and at 50~AU of HD~169142).
At the gap of HL~Tau, the dust filtration does not seem so effective becasue no significant dust pilling--up is seen at outer edge of the gap.

We also note that in both the observations of HL~Tau and HD~169142, the gaps we have focused on are only marginally resolved.  Therefore, the gap depth quoted in this paper should be regarded as upper limit.  Future high resolution observations by ALMA and/or TMT should put good constraints on the gap depth.  When complex morphologies are found in protoplanetary disks, simple analytic formulae should be useful for interpreting (at least a part of) such structures and derive physical quantities.

\acknowledgements
KDK thanks Tatsuya Takekoshi for helpful comments.
This paper makes use of the following ALMA data:
ADS/JAO.ALMA\#2011.0.00015.SV.  
ALMA is a partnership of ESO (representing its member states), NSF (USA)
and NINS (Japan), together with NRC (Canada) and NSC and ASIAA (Taiwan),  
in cooperation with the Republic of Chile. The Joint ALMA Observatory is 
operated by ESO, AUI/NRAO and NAOJ.
This work is partially supported by
JSPS KAKENHI Grant Numbers 23103004, 26103701, 26800106.
Numerical computations were carried out on Cray XC30 at Center for
Computational Astrophysics, National Astronomical Observatory of Japan.
This work is also greatly benefited from Workshop hosted by Institute of Low
Temperature Science, Hokkaido University 
"Recent Development in Studies of Protoplanetary Disks with ALMA".

\clearpage

\clearpage


\clearpage


%
%

\clearpage

\end{document}